\documentclass[aps,prl,reprint,superscriptaddress]{revtex4-1}

\usepackage{graphicx}
\usepackage{amsmath}
\usepackage{textcomp} 
\usepackage{xcolor}
\usepackage{ulem}
\usepackage{amssymb}
\usepackage{bm}

\begin{document}

\title{High frequency dynamics modulated by {collective magnetization reversal} in artificial spin ice}

\author{Matthias~B.~Jungfleisch}
\email{jungfleisch@anl.gov}
\affiliation{Materials Science Division, Argonne National Laboratory, Argonne, Illinois 60439, USA}

\author{Joseph~Sklenar}
\affiliation{Department of Physics and Frederick Seitz Materials Research Laboratory, University of Illinois at Urbana-Champaign, Urbana, Illinois 61801, USA}

\author{Junjia~Ding}

\affiliation{Materials Science Division, Argonne National Laboratory, Argonne, Illinois 60439, USA}

\author{Jungsik~Park}

\affiliation{Department of Physics and Frederick Seitz Materials Research Laboratory, University of Illinois at Urbana-Champaign, Urbana, Illinois 61801, USA}

\author{John~E.~Pearson}
\affiliation{Materials Science Division, Argonne National Laboratory, Argonne, Illinois 60439, USA}

\author{Valentine~Novosad}
\affiliation{Materials Science Division, Argonne National Laboratory, Argonne, Illinois 60439, USA}

\author{Peter~Schiffer}
\affiliation{Department of Physics and Frederick Seitz Materials Research Laboratory, University of Illinois at Urbana-Champaign, Urbana, Illinois 61801, USA}

\author{Axel~Hoffmann}
\affiliation{Materials Science Division, Argonne National Laboratory, Argonne, Illinois 60439, USA}

\date{\today}

\begin{abstract}

Spin-torque ferromagnetic resonance (ST-FMR) arises in heavy metal/ferromagnet heterostructures when an alternating charge current is passed through the bilayer stack. The methodology to detect the resonance is based on the anisotropic magnetoresistance, which is the change in the electrical resistance due to different orientations of the magnetization. In connected networks of ferromagnetic nanowires, known as artificial spin ice, the magnetoresistance is rather complex owing to the underlying collective behavior of the geometrically frustrated magnetic domain structure. Here, we demonstrate ST-FMR investigations in a square artificial spin-ice system and correlate our observations to magnetotransport measurements. The experimental findings are described using a simulation approach that highlights the importance of the correlated dynamics response of the magnetic system. Our results open the possibility of designing reconfigurable microwave oscillators and magnetoresistive devices based on connected networks of nanomagnets.

\end{abstract}

\maketitle

State-of-the-art nano-patterning enables the fabrication of networks of connected ferromagnetic nanowires, which serve as magnetic metamaterials. 
The class of these structures known as artificial spin ice (ASI) is designed to mimic the frustrated behavior of crystalline spin ice. ASI systems show degenerate ground states, complex magnetic ordering and collective behavior \cite{Heyderman_JPCM2013,Nisoli_RMP2013,Gilbert_Physics_Today}. ASI structures are not only model systems in which to study geometrical frustration, but have also significant technological potential as reconfigurable metamaterials and magnetic storage media \cite{Wang_Science_2016,Nat_Haldar_2016}. On the other hand, ASI can be considered as a reconfigurable magnonic crystal, 
which can serve as an integral part in magnonic circuits to tailor spin-wave properties by design \cite{Yu_JAP_2007,Wang_PRB_2005,Ding_JAP_2011}. 

Magnetoresistance (MR) and magnetization dynamics in nanostructures are two of the most important topics in contemporary studies of magnetism \cite{Hoffmann_PRAppl_2015}. In this regard, various aspects of MR in nano\-structures and ASI structures in particular have been reported. The first MR measurements in ASI systems indicated that anisotropic magnetoresistance (AMR) \cite{Mcguire_IEEE1975}, combined with nanowire domain reversal events, were responsible for field-dependent MR \cite{Tanaka_PRB_2006,Branford_Science_2012,Le_NewJPhys_2015}. Recently, angular- and field dependent magnetoresistive measurements and simulations of networks of connected Ni$_{81}$Fe$_{19}$ (Py) nanowires quantitatively revealed the importance of the vertex regions to the AMR \cite{Le_PRB2017,Jungsik_TBA}.  
Furthermore, complex dynamics in the high-frequency range have been observed in ASI. It was shown that the specific behavior of individual modes in the spectra can be correlated with the configuration of the magnetic moments in the nanowire legs \cite{Gliga_PRL2013,Jungfleisch_PRB_Ice_2016,Zhou_Adv2016,Bhat_PRB2016,Jungfleisch_APL2016,Iacocca_PRB2016,Li_JPD2016,Li_JAP2017,Sklenar_JAP2013,Ribeiro_JPC2016}. 

A big step forward towards the integration of such high-frequency devices in (spin-) electronics is the utilization of the spin-Hall effect in a heavy metal layer allowing for the conversion of electric charge currents into spin-polarized electron currents. In a bilayer of a ferromagnet and a heavy metal such as Pt or Pd, this spin current diffuses from the heavy metal to the ferromagnet where it exerts a torque on the magnetization leading to magnetization precession. If the ferromagnet is metallic, this precession results in a time-varying resistance in the ferromagnet on account of the AMR. Spin-torque ferromagnetic resonance (ST-FMR) combines these two effects to electrically drive and detect spin dynamics \cite{LiuPRL2011}. The main advantage of ST-FMR over conventional FMR is that it does not rely on an inductive detection technique which simplifies detecting resonant excitations in ferromagnets. Furthermore, the spin torque can assist the excitation of spin dynamics to induce unidirectional effects in the dynamic motion of the magnetization \cite{Sklenar_PRB2017}.  
ST-FMR was studied in a wide range of materials, compositions and structures, \cite{LiuPRL2011,Mellnik_Nat_2014,Jungfleisch_PRL_2016,Zhang_PRB_2015,Sklenar_PRB_2015,Zhang_APLMat_2016,Jungfleisch_PRB_Rashba_2016}. However, despite their interesting unconventional collective behavior and magnetotransport properties, magnonic crystals and ASI have not been explored by ST-FMR yet.

\begin{figure}[t]
\includegraphics[width=1\columnwidth]{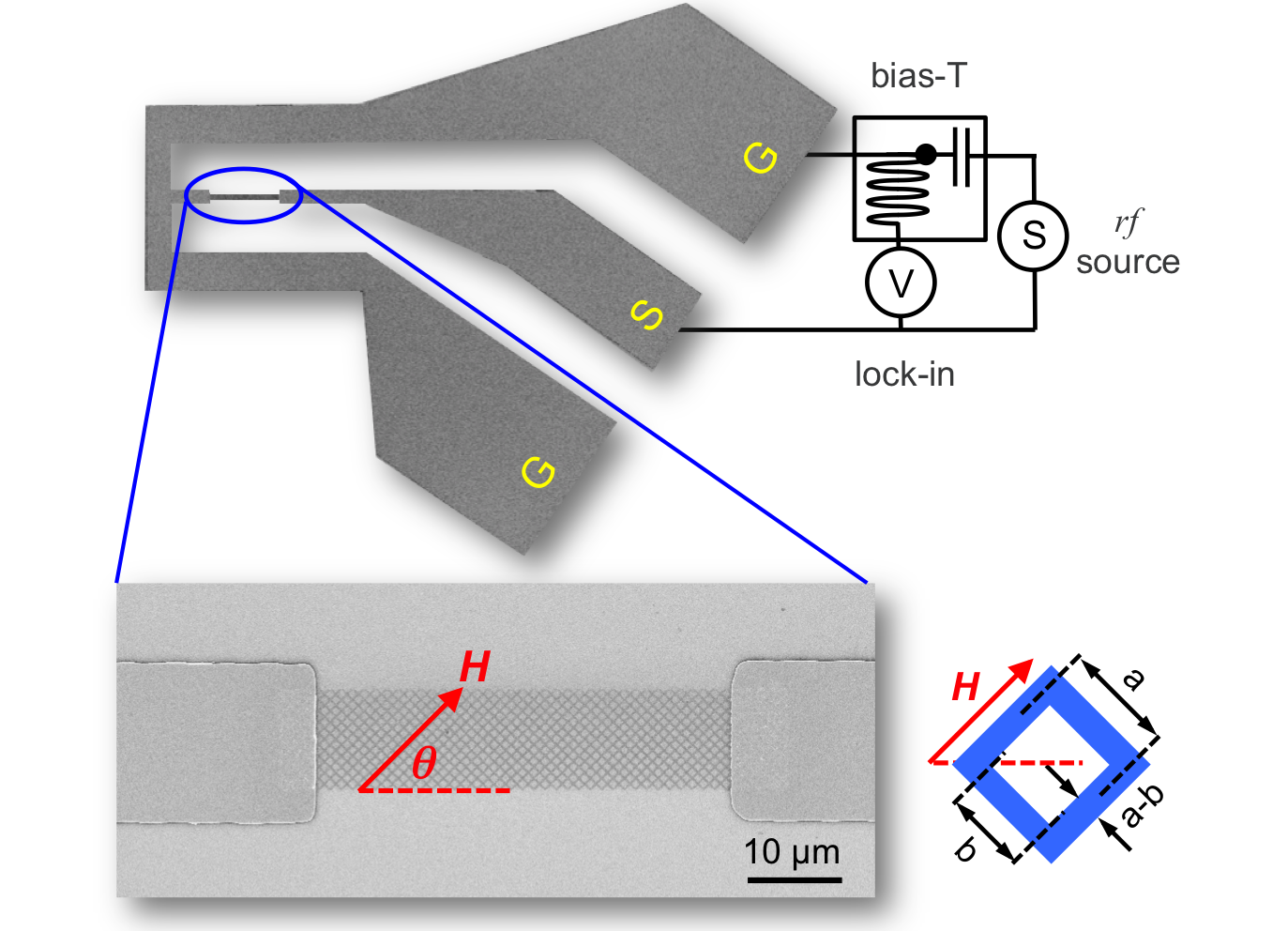}
\caption{\label{Fig1} (Color online) Experimental setup consisting of a shortened CPW made of Ti/Au with the square ASI made of Py/Pt integrated into the signal arm (S), see inset. A bias-T allows for simultaneous $rf$-signal transmission and 
voltage detection by lock-in technique (ST-FMR). A multimeter is used for the MR measurements (two-wire). 
The lattice constant is given by $a$, the hole width by $b$ and $c=a-b$ is the bar width. A magnified SEM image is shown in the supplementary material,  Fig.~S1.} 
\end{figure}

In this Letter, we study networks of permalloy/platinum (Ni$_{80}$Fe$_{20}$/Pt) nanowires arranged on a square lattice by ST-FMR. Complementary to these dynamic characterizations in the $rf$-frequency range, we carry out magnetotransport measurements and find distinct features in the longitudinal MR data, which correspond closely to features observed in the ST-FMR measurements. Correlating the experimental angular-dependent transport data with a micromagnetic-based transport model reveals that the sharp features in both the MR and the ST-FMR can be explained through the interaction of collective magnetization switching of the spin-ice lattice and AMR.  We find evidence that {the dynamic ST-FMR measurements are more sensitive to the occurrence of the switching of the nanowire legs triggered by external field changes than the magnetoresistance measurements.}
Our results have direct implications for magnetoresistive and spin-torque devices associated with complex magnetic nanostructures, {where reconfigurable magnonic and/or spin-ice states can be prepared by applying a specific charge current magnitude, magnetic field value and/or microwave power and frequency.}

The samples were fabricated in the following fashion: First, the square spin-ice structures of various dimensions were defined by electron beam lithography. 
15~nm-thick permalloy and 5~nm-thick Pt layers were deposited using magnetron sputtering at rates $< 0.7$~\AA/s without breaking the vacuum. 
The ASI lattices cover an area of approximately $75 \times 10~\mu$m$^2$ in total, and the lateral dimensions of each investigated lattice are {summarized in Tab.~\ref{table}.}
In a subsequent step, a shortened coplanar waveguide (CPW) made of Ti/Au (3 nm/120 nm) was fabricated by electron beam evaporation and photolithography. Figure~\ref{Fig1} shows a typical scanning electron microscopy image (SEM); the inset shows a zoomed-in view of the ASI structure and the dimensions of the lattice (magnified SEM image in the supplementary materials \footnote{For more information see supplementary material.}, Fig.~S1). 


\begin{table}[b]
\label{table} 
\caption{ \label{table} {Studied spin-ice lattices. $A, B, C$: Py thickness: 15~nm, Pt thickness: 5~nm. Parameters illustrated in Fig.~\ref{Fig1}.}}
\centering  

\begin{tabular}{c  c  c} 

\hline     \hline               
Lattice &  hole width $b$ (nm) & bar width $c$ (nm)\\ [0.5ex] 
\hline                     
$A$ & 620 & 252  \\ 
$B$ & 751 & 257 \\
$C$ & 497 & 254 \\     
\hline \hline
\end{tabular}

\end{table}

Figure~\ref{Fig1} illustrates the experimental setup. It consists of a shortened CPW with the square ASI structure integrated into the signal arm. For the ST-FMR measurements a bias-T is used to allow for a transmission of a high-frequency signal from a $rf$ source and simultaneous 
voltage detection by a lock-in amplifier (modulation frequency 3 kHz). The geometry of our ST-FMR experimental setup is limited to a fixed orientation (in-plane magnetic field parallel to one of the main axes of the lattice, $\theta \approx 45^\circ$, see Fig.~\ref{Fig1}). For the MR measurements, a two-wire method was used by connecting the CPW to a Keithley Nanovoltmeter (2182A). A current of 5~$\mu$A is provided by a Keithley current source (6221) 
The transport measurements were performed as a function of the in-plane angle $\theta$, see Fig.~\ref{Fig1}.  

\begin{figure}[t]
\includegraphics[width=1\columnwidth]{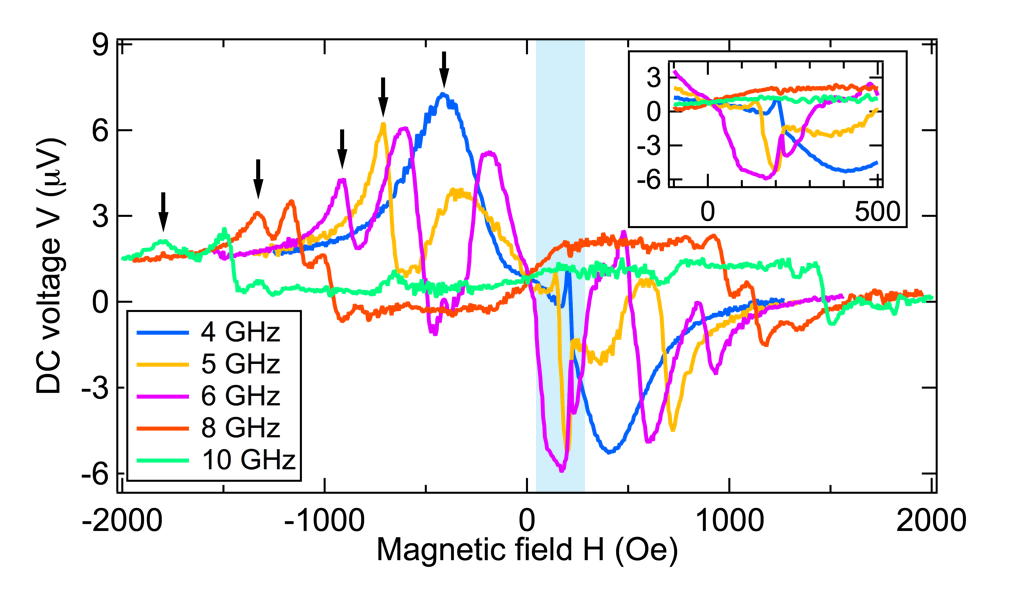}
\caption{\label{Fig2} (Color online) Frequency-dependent ST-FMR spectrum (microwave power: 6~dBm), here: lattice $A$. The {blue}-shaded area highlights the anomaly observed in the ST-FMR data. 
The inset shows the low field regime around the {blue}-shaded area on a magnified scale. {The resonance shifts to higher magnetic fields as the frequency is increased (black arrows). Figure~S2 shows the corresponding frequency-field dependence.}} 
\end{figure}

Figure~\ref{Fig2} shows an example of a ST-FMR spectrum taken on lattice $A$.  The results are consistent between the different lattices $A$, $B$, $C$. For the investigated ASI samples, a typical FMR spectrum is more complex than those observed from simpler micro- and nanostructures \cite{LiuPRL2011,Mellnik_Nat_2014,Jungfleisch_PRL_2016,Jungfleisch_Nano2017} and differ from high-frequency spectra of ASI structures \cite{Jungfleisch_PRB_Ice_2016,Jungfleisch_APL2016}. We also note that passing a $rf$ signal through the nanostructure results in 
inhomogeneous phase variation in the magnetization dynamics; this precludes a lineshape analysis and, thus, the determination of the spin-Hall angle. {However, it is yet possible to clearly identify the various modes using spatially-resolved dynamic micromagnetic simulations as will be discussed below.} 
 As is apparent from Fig.~\ref{Fig2}, the resonance shifts to higher magnetic fields as the frequency is increased (black arrows). 
Independent of the lattice parameters, the highest-lying mode (indicated by black arrows in Fig.~\ref{Fig2}) stiffens with increasing magnetic field, see Fig.~S2 in the supplementary material. 
Power-dependent ST-FMR measurements reveal a linear dependence of the resonance signal {at all tested frequencies}, Fig.~S3 in {the supplementary material shows exemplarily the results for $f=4$ GHz}.

%

\begin{figure}[t]
\includegraphics[width=1\columnwidth]{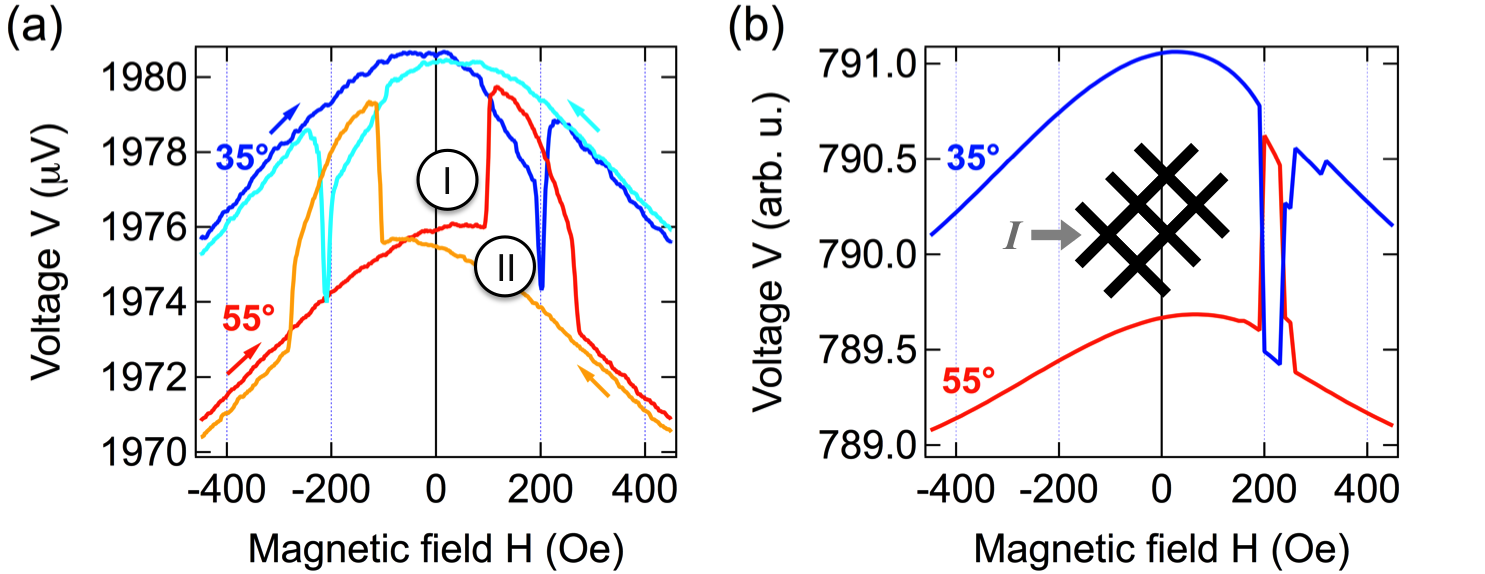}
\caption{\label{Fig3} (Color online) (a) Experimental MR data for in-plane angles $\theta = 35^\circ$ and $\theta = 55^\circ$, lattice $A$. Measurement configuration as shown in Fig.~\ref{Fig1} and sketched in the inset of (b); $I$ is the charge current. The negative field sweeps and positive field sweeps (small arrows) are symmetric under field reversal. The field range (positive sweep direction) {labeled (I) and (II) indicate fields before and after a collective change in the magnetization (`\textit{avalanche}') is triggered, respectively}. (b) Corresponding simulated MR (only positive sweep shown).} 
\end{figure}

Besides the expected resonances in the ST-FMR data, a surprising feature is observed in the spectra: Independent of the excitation power, we find an anomaly in the voltage spectra of all lattices. This anomaly is characterized by a distinct change in the detected voltage, see {blue} shaded area in Fig.~\ref{Fig2} and Fig.~S3(a) in the supplementary material. The inset in Fig.~\ref{Fig2} shows the sharp feature at low magnetic fields on a magnified scale. The negative field sweeps and positive field sweeps are symmetric under field reversal (not shown here). Dependent on the lattice parameter and excitation frequency, this jump in the voltage is more or less pronounced. Furthermore, the field at which the jump occurs, varies slightly for different driving frequencies. Naturally, one might ask if our observation is due to the complex collective behavior of the studied magnetic nanostructure? We address this question by exploring magnetotransport measurements in the studied nanostructures and correlating them with a micromagnetic-based transport model that gives us a microscopic understanding of our results. In addition, we carry out detailed micromagnetic simulations to understand the spatially-resolved dynamics.

The measured MR at two different in-plane angles $\theta = 35^\circ$ and $\theta = 55^\circ$ is depicted in Fig.~\ref{Fig3}(a) (up and down sweep indicated by arrows next to the traces). 
The traces for the two measurement angles are significantly different. The $35^\circ$ curve starts off at a higher voltage than the $55^\circ$ curve (at given magnetic field), has a maximum at zero-field and drops significantly at a small magnetic field at $ 200$~Oe before it recovers and then slowly decreases with increasing field. In contrast, the $55^\circ$ trace slowly increases while the field is swept up, saturates around zero-field and then jumps up at $200$~Oe before it decreases again. The negative field and positive field sweeps (small arrows) are symmetric under field reversal.

\begin{figure}[t]
\includegraphics[width=1\columnwidth]{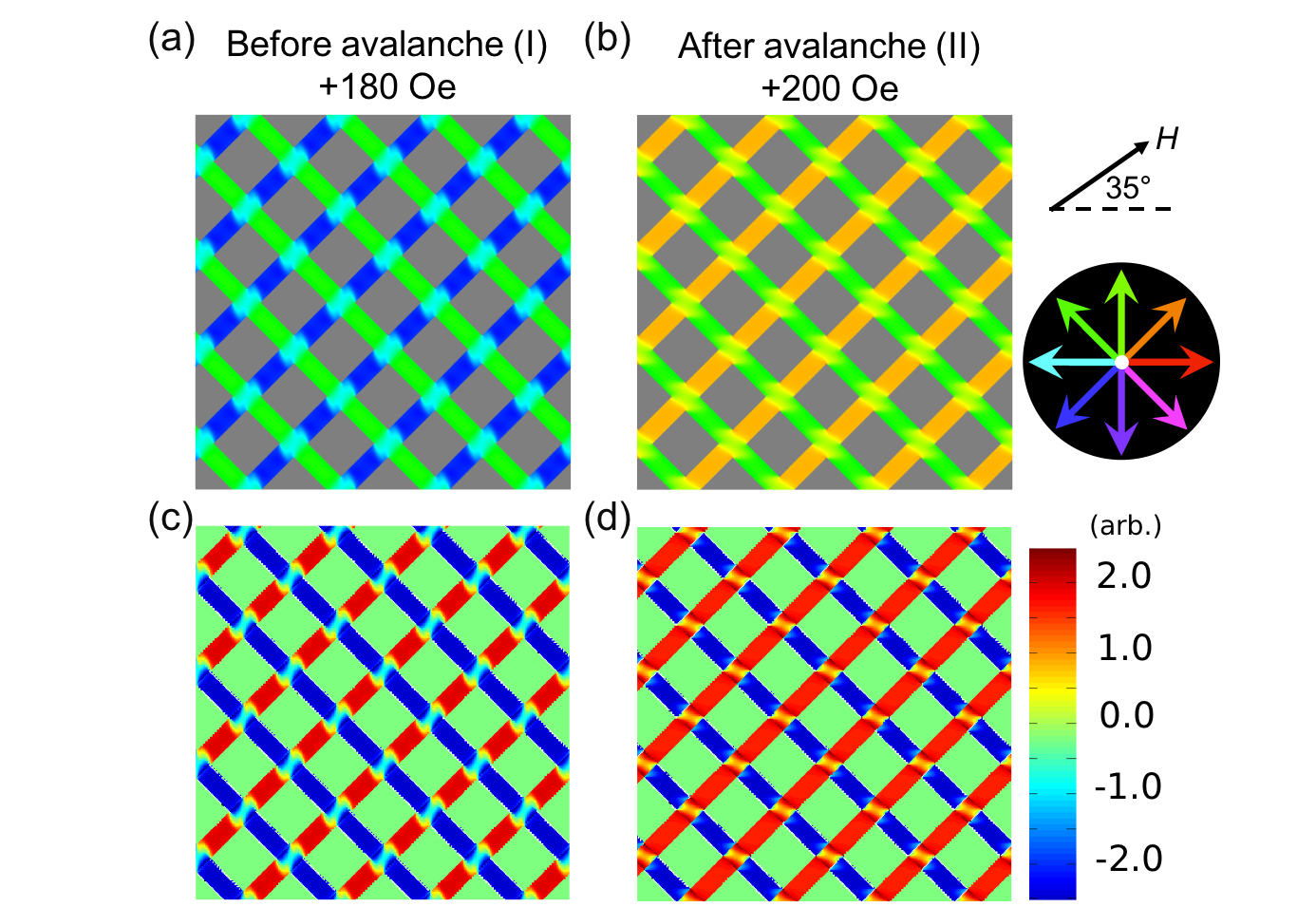}
\caption{\label{Fig4} (Color online) Simulated magnetization configuration (a,b) and $y$-component of the electric field {(c,d)}, respectively. Results of the micromagnetic simulations at $\theta = 35^\circ$ before (a) and after (b) the collective switching occurs; denoted as (I) and (II) in Fig.~\ref{Fig3}(a). (c) and (d) show the electric field maps of the respective states. The sample is first magnetized in the negative field direction and then swept in the positive direction. The sharp features in the MR, Fig.~\ref{Fig3}, is mainly determined by the vertex region: they realign from horizontal to vertical direction when the avalanche is triggered. This switching in the vertex region causes changes in the AMR {as is apparent from the change in the polarity of the $y$-component of the electric field in (c) and (d).  We note that this effect is also seen in the $x$-component of the electric field as shown in the supplementary material, Fig.~S5.}} 
\end{figure}

To better understand the underlying mechanisms in the observed MR behavior of our nanostructures, we use a combination of micromagnetic simulations and the phenomenology of AMR, following Ref.~[\onlinecite{Le_PRB2017}]. The magnetization profiles were obtained using Mumax3 \cite{mumax3}. In order to simulate the experimentally acquired field-dependent resistance traces the micromagnetic field maps 
\footnote{A grid of $1024\times1024$ micromagnetic cells was used. The volume of each cell was $10\times10\times15$~nm$^3$.  A grid of this size with the given pixel volume leads to approximately 14 lattice constant to span the diagonal of the simulation space. The micromagnetic states were obtained by sweeping the magnetic field at a fixed angle $\theta$ from $-600$~Oe to $+600$~Oe with a step size of 10 Oe.  The resulting static equilibrium configuration at each step was found by using an energy minimization function.} 
are converted to electric field maps that take into account first order changes in the electric field due to AMR \cite{Mcguire_IEEE1975}. {The electric field associated with anisotropic magnetoresistance is given by:}
\begin{equation}
{\bm{E}=\rho_0 \bm{J}+\Delta \rho(\bm{\hat{m}}\cdot\bm{J})\bm{\hat{m}},}  \label{AMR_E_field}
\end{equation}
{where $\bm{J}$ is the electric current vector, $\bm{\hat{m}}$ is the unit vector of the magnetization, $\rho_0$ is the isotropic resistivity, and $\Delta \rho$ is the anisotropic magnetoresistivity.} 
The first order electric field can be integrated along paths that mimic the experimental configuration to then simulate the observed MR \cite{Le_PRB2017}. We assume that half of the current $I/2$ flows in either of the nanowire legs and the full current $I$ flows in the vertex, see Fig.~S4 (supplementary material). 
A remarkably good agreement between experimentally observed [Fig.~\ref{Fig3}(a)] and modeled MR [Fig.~\ref{Fig3}(b)] is found. Although the model only requires the lattice parameters as input, it captures the main experimental findings. 
Additional MR measurements on square networks made of a single layer permalloy confirm our findings and highlight the generality of our model{, see Fig. S6 in supplementary material.}

Snapshots of our simulations at characteristic field values at an in-plane angle of $35^\circ$ are shown in Fig.~\ref{Fig4}  [denoted as (I) and (II) in Fig.~\ref{Fig3}(a)]. Figures~\ref{Fig4}(a,c) show the magnetization and respective electric field maps right before the sharp feature in the magentoresistance is observed [$H< 200$~Oe, labeled as (I)]. Respectively, Fig.~\ref{Fig4}(b) and (d) illustrate the maps right after the sharp features in the MR have been observed [$H\approx 200$~Oe, labeled as (II)]. The system is initially magnetized at a negative field and  then swept in positive field direction. The micromagnetic simulations reveal that sharp features in the MR data occur when a rapidly evolving collective change in the magnetization (`\textit{avalanche}') happens [(I) and (II) in Fig.~\ref{Fig3}(a)]. {The avalanches in our artificial spin ice are triggered by the external magnetic field.} The distinct signature in the MR is mainly determined by the vertex region: 
The moments in the vertex are initially aligned horizontally in the negative field direction and reorient to the vertical direction after the avalanche is triggered. 
This switching from a horizontal to a vertical alignment causes a change in the AMR as is obvious from the corresponding electric field maps shown in Fig.~\ref{Fig4}(c) and (d) 
\cite{Le_PRB2017}. At the vertices, the polarity of the $y$-component of the electric field switches states, while the electric field in the remaining portion of the network remains unchanged. {We note that this effect is also seen in the $x$-component of the electric field as shown in supplementary material, Fig.~S5.}
As the magnetic field increases, the moments in the vertex region re-align horizontally in the positive field direction and the MR goes back to approximately the same value as before. 



The simulation results indicate that the MR data for an in-plane measurement angle of $35^\circ$ differs strongly from the results obtained at $55^\circ$ since the magnetic moments in the vertex region are orthogonal to each other. This is why the MR drops when the islands switch at $35^\circ$, and increases at $55^\circ$: The electric current and the magnetization are perpendicular to each other at $35^\circ$, and they are parallel to each other in the $55^\circ$ configuration (see 
Fig.~S7 in supplementary material).

Our results clearly reproduce the previous experimental MR results \cite{Le_PRB2017,Jungsik_TBA} that the vertex regions in spin-ice structures strongly influence experimentally measured MR. 
Strikingly, the occurrence of the sharp features in the transport correlates well with the appearance of corresponding signatures in the dynamic ST-FMR data at the same magnetic fields. This leads to the question: Do the observed features in the dynamic spectra stem from an AMR mixing mechanism related to a dynamic mode of the system that appears at/during the onset of an avalanche or from the collective switching itself? 

\begin{figure*}[t]
\includegraphics[width=1.55\columnwidth]{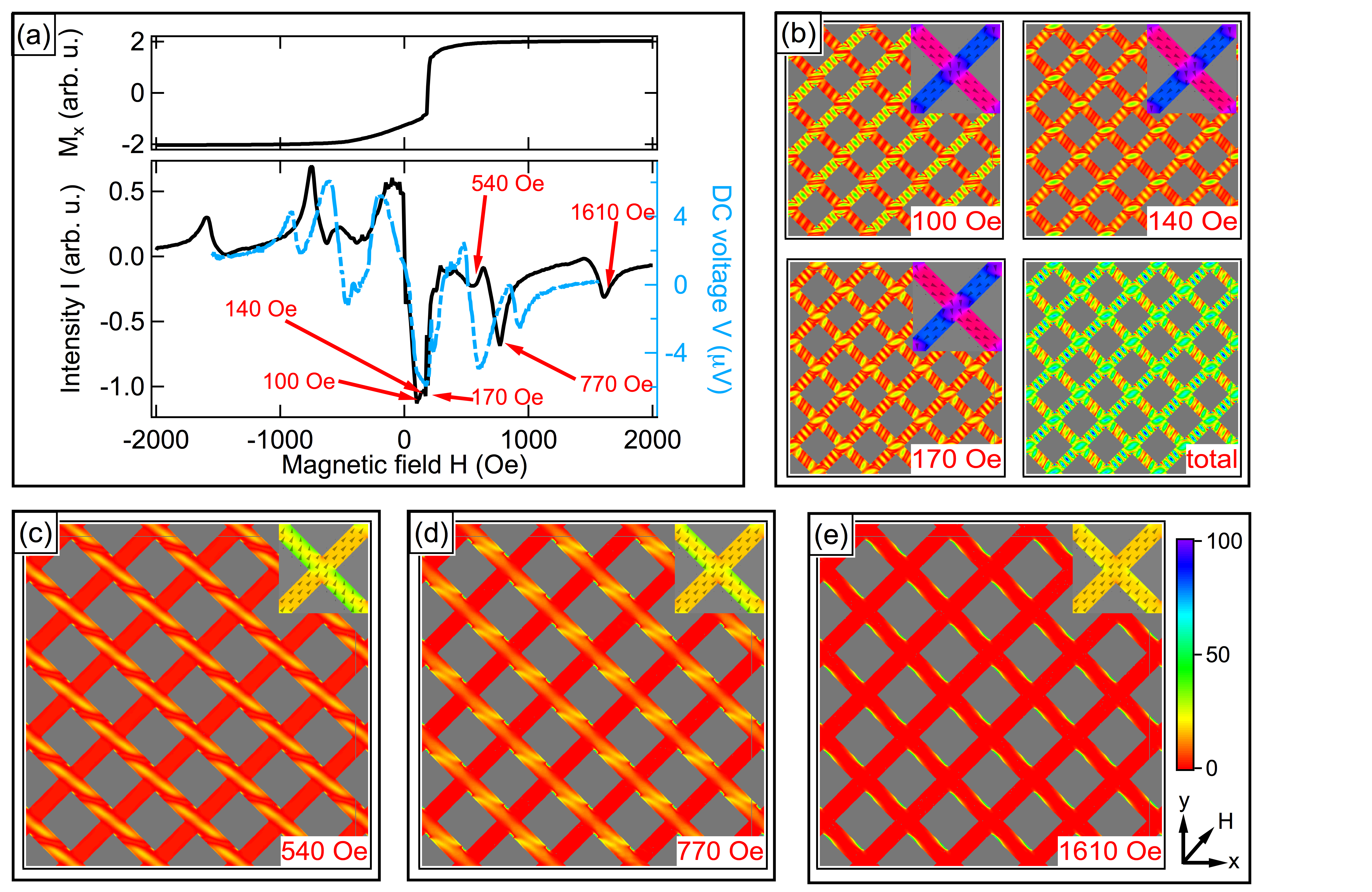}
\caption{\label{Fig5} (Color online) {Results of dynamic micromagnetic simulations. (a) Top panel: Simulated magnetization in $x$-direction as a function of the external field, which is applied at $50^\circ$ in the simulations. The coercive field is about 190 Oe. Bottom panel: Predicted resonance modes (black, straight line) and their comparison to experimental spectrum (blue, dashed line) at 6 GHz. A reasonable agreement between experiment and simulation is found. Please note that the $m_\mathrm{z}$-component of the dynamic magnetization has been used to represent the magnetization precession. (b-e) Spatially-resolved maps of the dynamic response at a given magnetic field. The color scale represents the dynamic $m_\mathrm{z}$-component of the precession at 6 GHz and illustrates where the particular modes at a given magnetic field are located. The insets show the corresponding static magnetization configuration for comparison. 2D maps of the precession are shown for the fields indicated in (a): (b) Reversal region, (c) bulk mode, (d) higher order mode and (e) edge mode. The static magnetization configuration in the reversal regime, (b), only changes slightly, while the dynamic maps show a big change when the field is swept.}} 
\end{figure*}

To address this question, we carried out dynamic micromagnetic simulations 
\footnote{The spatial characteristics of different modes were simulated using time-dependent
micromagnetic simulations using Mumax3 \cite{mumax3} and analyzed by spatially and frequency-resolved fast Fourier transform imaging at various external magnetic field values.}. {The results are summarized in Fig.~\ref{Fig5}. Figure~\ref{Fig5}(a) shows a simulated $M/H$ loop (top panel) and the predicted resonance modes and their comparison to the experimental observations. Exemplarily the results for a driving frequency $f=6$~GHz is shown (bottom panel). The in-plane magnetic field angle in the simulations was chosen so as to resemble the experimental configuration, but at the same time to break the lattice symmetry (50$^\circ$). 
The $y$-scale of the simulated spectra is the dynamic response of the $m_\mathrm{z}$-component (intensity of precession $\propto m_\mathrm{z}^2$) to a constant sinusoidal driving frequency of 6 GHz as a function of the magnetic field. A low intensity means that there is no resonance at 6 GHz at a given magnetic field. A strong intensity means a large precession amplitude at 6 GHz at a given magnetic field.
A good agreement between simulation and experiment is found. In particular, the field asymmetry in the range between +100 Oe and +170 Oe found in the experiment is also observed in the dynamic micromagnetic simulations. This field range also agrees very well with the reversal regime found in the magnetoresistance measurements (Fig.~\ref{Fig3}). Meanwhile the resonances detected at +600 Oe and +930 Oe in the experiment, are found at +540 Oe and +770 Oe in the simulations. The difference between micromagnetics and experiment may be attributed to details of imperfections in our simulations; e.g., simulations occur at 0 K and do not account for any lithographic imperfections. Furthermore, the simulation only considers the response of a limited number of unit cells with periodic boundaries due to computational limitations, and assumes that the roughness is uniform across the entire sample. The predicted mode at +1610 Oe was not detected in the experiment; as we will discuss below, this mode is an edge mode. Typically edge modes occur in simulations, in which a perfect structure without any imperfections is assumed. Clearly, this assumption is not fulfilled in real spin-ice networks which may be the reason why we do not observe this mode in the experiment. Please note that the polarity of the simulated intensity was manually mirrored for negative fields to better match the experimental findings (which are detected based on ST-FMR). The reason is that Mumax3 simulates the resonances, not ST-FMR directly which changes sign due to the spin-Hall effect.}

{In the following, we discuss the two-dimensional maps of the dynamic magnetization at the resonances indicated in Fig.~\ref{Fig5}(a). These maps illustrate where in the spin-ice lattice the intensity is the largest (small intensity: red, large intensity: blue).  We also show the corresponding static magnetization configuration of the vertex regions as insets. 
Figure~\ref{Fig5}(b) shows the low-field range at +100 Oe, +140 Oe and +170 Oe, as well as the total magnetization dynamics in the field range, which is the sum of +100 Oe, +140 Oe and +170 Oe. The dynamics in this field range is quite uniform. 
From a comparison of the static and dynamic maps in the low field range we conclude that the dynamic signal is much more sensitive to external field changes, which implies that the dynamic measurements are more sensitive to the occurrence of avalanches than the magnetoresistance measurements since they rely only on the static magnetization.}
The sharp features due to avalanche formation could not be observed in similar spin-pumping experiments \cite{Jungfleisch_APL2016}. This might indicate that an inhomogeneous phase of the microwave signal, out-of-plane Oersted fields and/or spin torques are required to couple to {this low-field mode. This mode is most pronounced at 6 GHz in the experiment as well as in the simulations. It is non-existent in the experiment at higher frequencies and hardly observable in the corresponding simulations. At lower frequencies this low-field mode is narrower in field range and less intense than at 6 GHz. This experimental observation is confirmed in the simulations; not shown here.}

{Figure~\ref{Fig5}(c-e) show the simulated two-dimensional maps of the dynamic response at +540 Oe, +770 Oe and +1610 Oe. The mode at +540 Oe is identified as a bulk-like mode, Fig.~\ref{Fig5}(c), while the resonance at +770 Oe corresponds to a higher order mode, Fig.~\ref{Fig5}(d), and the resonance at +1610 Oe is an edge mode, Fig.~\ref{Fig5}(e).}

We emphasize that the ST-FMR rectification is given by a time-averaged mixing of the microwave current with the resistance, $V = \langle R I\rangle$. Therefore, it is likely that two components contribute to the ST-FMR signal: (1) a static (heterodyne) magnetoresistance change 
and (2) a dynamic (homodyne) anisotropic resistance due to the formation of 
{a change of the magnetization configuration}, occurring due to the switching, that is susceptible to the microwave/spin-torque drive.


In summary, we demonstrated that the collective magnetization behavior in an ASI strongly affects the dynamic ST-FMR spectra, as well as the magnetoresistive behavior in those structures. We provide a microscopic picture of this unexpected response by means of micromagnetic simulations. The sharp features observed in the experimental data occur due to a sudden change in the magnetization configuration when the field is swept. We show that the angular-dependent alignment of the vertex region strongly affects the resistance and even more importantly the resonance condition leading to a significantly different ST-FMR response. {Our findings clearly demonstrate the possibility to \textit{read out} collective switching processes in ASI by transport, as well as high-frequency dynamics. The observation of a spatially-confined magnonic modes in ASI is a first step towards the realization of microwave oscillators in connected networks of ferromagnetic nanowires.} Given the geometric freedom enabled by modern lithography techniques, our results open the possibility of designing innovative reconfigurable microwave oscillators and magnetoresistive devices based on connected ferromagnetic networks which might also provide desirable functionalities, e.g., for neuromorphic computing \cite{Grollier_IEEE_2016}.

\begin{acknowledgments}
Work at Argonne including experiment design, sample fabrication and characterization, ST-FMR and magnetoresistance measurements, mircomagnetic simulations, data analysis, and manuscript preparation, was supported by the U.S. Department of Energy, Office of Science, Materials Science and Engineering Division. Lithography was carried out at the Center for Nanoscale Materials, an Office of Science user facility, which is supported by DOE, Office of Science, Basic Energy Science under Contract No. DE-AC02-06CH11357. Work at UIUC including sample fabrication, magnetoresistance measurements, micromagnetic simulations and data analysis was supported by the U.S. Department of Energy, Office of Basic Energy Sciences, Materials Sciences and Engineering Division under Grant No. DE-SC0010778.
\end{acknowledgments}
%

\end{document}